\documentstyle[epsfig,prl,aps,twocolumn]{revtex}

\tighten
\begin{document}
\bibliographystyle{/usr/share/texmf/tex/latex/revtex/prsty}
\draft

\title{Expansion of an interacting  Fermi gas}
\author{$^{1,2}$C. Menotti, $^1$P. Pedri and $^1$S. Stringari}
\address{$^1$ Dipartimento di Fisica, Universit\`{a} di Trento and BEC-INFM,}
\address{I-38050 Povo, Italy}
\address{$^2$ Dipartimento di Matematica e Fisica,
Universit\`{a} Cattolica del Sacro Cuore,}
\address{I-25121 Brescia, Italy}

\date{\today}

\maketitle

\begin{abstract}
\noindent
We  study the expansion of a dilute ultracold sample of fermions 
initially trapped in a anisotropic harmonic trap. The expansion of 
the cloud provides valuable information about the state of the system 
and the role of interactions. In particular the time evolution of the 
deformation of the expanding cloud behaves quite differently depending 
on whether the system is in the normal or in the superfluid phase. 
For the superfluid phase, we predict an inversion of the deformation of 
the sample, similarly to what happens with Bose-Einstein condensates. 
Viceversa, in the normal phase, the inversion of the aspect ratio is 
never achieved, if  the mean field interaction is attractive and 
collisions are negligible.\\ 
\end{abstract}

Since the first experiments on trapped Bose-Einstein condensates the 
imaging of the expanding cloud, following the sudden switching off of 
the confining potential, has provided crucial information on the novel
features exhibited by  atomic gases in conditions of quantum degeneracy. These
include, in particular, the bimodal structure of an expanding Bose gas at
finite temperature and the anisotropy of the asymptotic profile of the 
condensate \cite{jila95,mit95}.
In the Thomas-Fermi limit, where the Gross-Pitaevskii equations  coincide with
the hydrodynamic theory of superfluids, the anisotropy of the expanded gas
reflects   the anisotropy of the pressure force which is stronger in the 
direction of tighter confinement \cite{rmp}. The  predictions of the 
hydrodynamic  equations and of the consequent scaling behaviour exhibited
by Bose-Einstein condensates during the expansion
have been investigated by  several authors 
(\cite{kagan},\cite{castin},\cite{dalfovo}), 
providing excellent agreement with experiments  
\cite{mit1,rempe} and pointing out the difference with respect to the
expansion of a non condensed gas. In the latter 
case, in the collisionless regime,  the density profile approaches  
an isotropic shape, independent of the initial 
deformation of the gas.
 
In this letter we  study the problem of the expansion of an ultracold
sample of fermions initially trapped in an anisotropic harmonic trap. 
We will show that also in the case of fermions the expansion of the gas 
provides valuable information about the state of the system and the role
of interactions. 
We will  consider a gas of atoms interacting with attractive forces.
This is a natural requirement for the realization of  Cooper-pairs 
and hence for the achievement of the superfluid phase  \cite{stoof}. 
Such interactions are naturally present 
in some fermionc species like  $^6$Li and can otherwise be obtained
by changing the scattering length profiting of the presence of a 
Feshbach resonance.

The description of the expansion of a cold fermionic gas in the normal and 
superfluid phase requires different theoretical approaches. For the 
normal phase we use the formalism of the Landau-Vlasov equations, while  in the 
superfluid phase we study the expansion using  the hydrodynamic  
theory of superfluids.

We consider the case of two different fermionic
states, hereafter called  1 and 2, initially confined 
 in a harmonic trap. We assume that the two species  are present
in the same amount and feel the same trapping potential,
so that the  densities of the two species 
are equal at equilibrium as well as during the 
expansion: $n_1({\bf r},t)=n_2({\bf r},t)=n({\bf r},t)/2$. The trapping potential
 will be chosen of 
 harmonic type
\begin{eqnarray}
V_{ho} =\frac{1}{2}m(\omega_{\perp}^2 x^2+\omega_{\perp}^2y^2+\omega_z^2z^2),
\label{Vho}
\end{eqnarray}
describing a cilyndrically simmetric trap with deformation 
$\lambda=\omega_z/\omega_{\perp}$. 
The interaction  between the two fermionic species
is fixed by the coupling constant
$g=4 \pi \hbar^2 a /m $, where $a$ is the $s$-wave scattering length. 

In this paper we will use the equation of state
\begin{eqnarray}
\mu_{\ell e}(n)=  {\hbar^2\over 2m} (3\pi^2 n)^{2/3} + \frac{1}{2} g n,
\label{mugg}
\end{eqnarray}
to describe the uniform phase of the gas 
where the first term is the kinetic energy  evaluated at zero temperaure
while the second one is the interaction energy evaluated in the mean field 
approximation.
Equation (\ref{mugg}) neglects the effects of correlations which become 
important for large values of the scattering length and affect in a 
different way the equation of state of the normal and superfluid phase.
The formalism developed in this paper can be easily generalized to include 
a more accurate description of the equation of state. It is however worth 
pointing out that, even using the same equation of state for the normal 
and superfluid phases, the expansion of the gas behaves quite 
differently in the two cases being described by different
kinetic equations.
In the presence of the external potential (\ref{Vho}),
the equilibrium condition in
the  local density approximation is determined by the equation

\begin{equation}
\mu_{\ell e}(n)+V_{ho}({\bf r})=\mu
\label{tf1}
\end{equation}  
where $\mu$ is the chemical potential of the sample fixed by the 
normalization condition.  

\begin{center}
\begin{figure}
\includegraphics[width=0.95\linewidth]{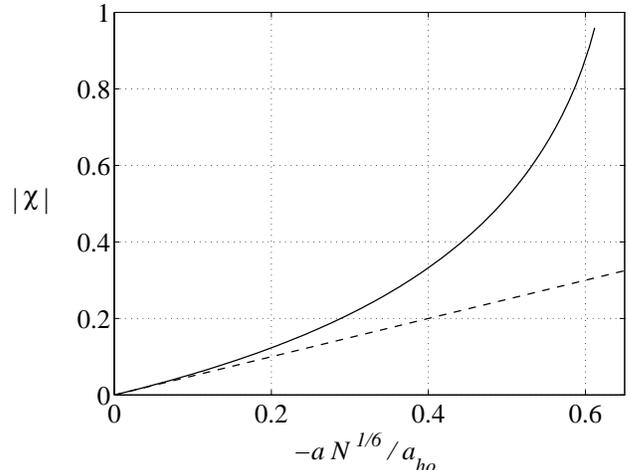}
\caption{Ratio $\chi=E_{int}/E_{ho}$ as a function of the universal parameter
$a N^{1/6} / a_{ho}$ calculated with the mean field functional (\ref{E})
up to the collapse point $a N^{1/6} / a_{ho}=-0.61$ (full line); 
the dashed line is the linear prediction (\ref{vichi}).}
\label{fig_beyond}
\end{figure}
\end{center}

The relevant parameter characterizing the interaction in the  fermionic system is
the ratio 
\begin{equation}
\chi = {E_{int} \over E_{ho}}
\label{chi}
\end{equation}
between the interaction energy

\begin{eqnarray}
E_{int} = \frac{g}{4}  \int n^2({\bf r}) d^3r
\label{Eint1}
\end{eqnarray}
and the  oscillator energy  

\begin{eqnarray}
E_{ho} &=&  \int V_{ho}({\bf r}) n({\bf r}) d^3 r.
\label{Eho1}
\end{eqnarray}
In the perturbative regime the integrals (\ref{Eint1},\ref{Eho1}) can be 
evaluated using the non interacting density profile which, in
Thomas-Femi approximation (\ref{tf1}), takes the simple form 

\begin{equation}
n({\bf r}) = \frac{1}{3 \pi^2} \left( \frac{2m}{\hbar^2} \right)^{3/2}
\left[ \mu -V_{ho}({\bf r}) \right]^{3/2}.
\label{densityIG}
\end{equation}
After integration of (\ref{Eint1}) and (\ref{Eho1}) one finds 
\cite{vichi}

\begin{eqnarray}
\frac{E_{int}}{E_{ho}} = 0.5 \frac{N^{1/6} a }{a_{ho}} = 0.3 a k_F,
\label{vichi}
\end{eqnarray}
where $k_F= (3\pi^2 n(0))^{1/3}$ is the Fermi momentum evaluated 
at the central value of the density.
In order to go beyond the perturbative regime, one determines 
numerically the ground state density by minimizing the  energy of 
the system 

\begin{eqnarray}
E &=&  \frac{3(3\pi^2)^{2/3}}{5} 
 \frac{\hbar^2}{2m}  \int n^{5/3}({\bf r}) d^3r + \nonumber \\
&+& \int V_{ho}({\bf r}) n({\bf r}) d^3 r +
 \frac{g}{4}  \int n^2({\bf r}) d^3r,
 \label{E}
\end{eqnarray}
One easily finds that the equilibrium value of the ratio 
$E_{int}/E_{ho}$ depends on the dimensionless combination $aN^{1/6}/a_{ho}$
also in the non perturbative regime \cite{stoof},
as reported in Fig.\ref{fig_beyond}. 
Using  (\ref{E}) one predicts that the compressibility of the gas  becomes 
negative in the center of the trap if 
$|a| N^{1/6}/a_{ho} > 0.61$.
For large values of $|a|$ the resulting predictions  
should be however taken with care since  the functional  (\ref{E}) 
ignores correlation effects beyond  mean field. 

Let us now discuss the expansion of a fermionic sample trapped in an elongated
harmonic trap ($\lambda<1$). We describe first the expansion
of the normal fluid and afterwards the one of the superfluid.
In the ideal case, using the semiclassical description, 
one finds that  the ratio of the square radii evolves according to 
the classical law

\begin{equation}
\frac{\langle r_{\perp}^2\rangle}{\langle z^2\rangle}=
\frac{1+\omega_{\perp}^2t^2}{1+\omega_z^2t^2}
\frac{\omega_z^2}{\omega_{\perp}^2},
\label{ideal}
\end{equation}
The ratio (\ref{ideal}) approaches unity for large times, reflecting the isotropy of the 
momentum distribution. This result ignores the  effects of
collisions which are however expected to play a minor role at low temperature, 
due to Pauli blocking, unless the scattering length is very large.

In order to take  into account the effects of the interactions,  we consider
the mean field description based on the  Landau-Vlasov equation

\begin{eqnarray}
\frac{\partial f}{\partial t} +
{\bf v} \cdot \frac{\partial f}{\partial {\bf r}} -
\frac{\partial V_{ho}}{\partial {\bf r}} \cdot
\frac{\partial f}{\partial {\bf v}}  -
\frac{1}{2} \frac{g}{m} \frac{\partial n}{\partial {\bf r}} \cdot
\frac{\partial f}{\partial {\bf v}} =0,
\label{BV}
\end{eqnarray}
where $f({\bf r},{\bf v},t)$ is the distribution function, $n=\int f d^3v$ 
is the atomic density. Eq.(\ref{BV}) describes the dynamics of a normal
weakly interacting gas in the collisionless regime.

An approximate solution of Eq.(\ref{BV}), 
is obtained by making a scaling ansatz for the distribution function
\begin{eqnarray}
f({\bf r},{\bf v},t)=f_0({\bf {\tilde r}}(t),{\bf {\tilde v}}(t)),
\label{scaling}
\end{eqnarray}
where $f_0$ is the equilibrium distribution,
${\tilde r}_i(t)=r_i / b_i$ and 
${\tilde v}_i(t)=b_i v_i - {\dot b}_i r_i$.
Under the scaling assumption the velocity field
${\bf u}({\bf r},t) = \int {\bf v} f d^3 v /n$ takes the simple form
$u_i={\dot b_i} r_i/b_i$. This ansatz has been recently used by Gu\'ery-Odelin
\cite{david} to investigate the effect of the interaction on the collective 
oscillation of a classical gas in the collisionless regime.

 The equations for the scaling parameters $b_i$ can be obtained by
multiplying (\ref{BV}) by ${\tilde r}_i$ and ${\tilde v}_i$ and integrating in 
phase space. Making use of the equilibrium properties of the distribution function, 
after some straigthforward algebra one finds

\begin{eqnarray}
\ddot{b}_i + \omega_i^2 b_i -  \frac{\omega_i^2}{b_i^3} +
\frac{3}{2} \chi
\omega_i^2  \left( \frac{1}{b_i^3} - \frac{1}{b_i \prod_j b_j} \right)
=0,
\label{intnormal}
\end{eqnarray}
where $\chi$ is the ratio (\ref{chi}) evaluated at equilibrium. 
 The second term in (\ref{intnormal})
describes the restoring force of the external oscillator  potential, 
the third one originates from the kinetic energy, while the last term, 
linear in $\chi$, accounts for the effects of  mean field interaction. 

An immediate application of  Eq.(\ref{intnormal}) concerns the study of the oscillations of the gas. By linearizing
the equations around  equilibrium  ($b_i=1$) one finds, in the presence
of isotropic harmonic trapping ($\omega_{\perp}=\omega_z=\omega_0$), the result:

%
\begin{eqnarray}
\omega_M = 2 \omega_0 \sqrt{1+3\chi/8}, \;\;
\omega_Q = 2 \omega_0 \sqrt{1-3\chi/4}, 
\label{omega_vichi}
\end{eqnarray}
for the frequencies of the monopole and quadrupole oscillations 
which coincide with the results already derived  in \cite{vichi}
using a sum-rule approach. 

\begin{center}
\begin{figure}
\includegraphics[width=0.95\linewidth]{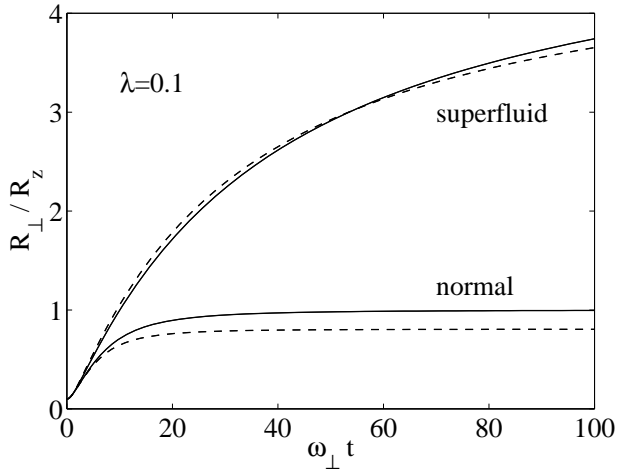}
\caption{Aspect ratio as a function of time for the expansion of the
normal (lower curves) and superfluid phase (upper curves) for $\lambda=0.1$
and two different values  of the parameter $\chi$: 
$\chi=0$ (full line) and $\chi=-0.4$ (dashed line).}
\label{fig1}
\end{figure}
\end{center}

The equations describing the expansion are obtained by suddenly removing 
the second 
term of Eq.(\ref{intnormal}), originating from  the trapping potential.
In the study of the expansion we are interested  in the case of 
anisotropic trapping. In particular we will consider
the case of cigar shaped traps. 
For high deformations   
($\lambda =\omega_z/\omega_{\perp} \ll 1$) Eq.(\ref{intnormal}) yields the asymptotic result 

\begin{eqnarray}
b_z^2 &\rightarrow& \omega_z^2 (1-{3\over 2}\chi) t^2,\\
b_{\perp}^2 &\rightarrow& \omega_{\perp}^2  t^2,
\end{eqnarray}
showing that the aspect ratio
\begin{eqnarray}
\frac{R_{\perp}(t)}{R_z(t)} \rightarrow \frac{1}{\sqrt{1-3\chi/2}} 
\label{ratio}
\end{eqnarray}
 approaches a value smaller than 1 if the interaction is attractive
($\chi <0$). In Eq.(\ref{ratio}), $R_{\perp}$ and $R_z$ are the radii 
where the atomic density vanishes (Thomas-Fermi radii).

\begin{center}
\begin{figure}
\includegraphics[width=0.95\linewidth]{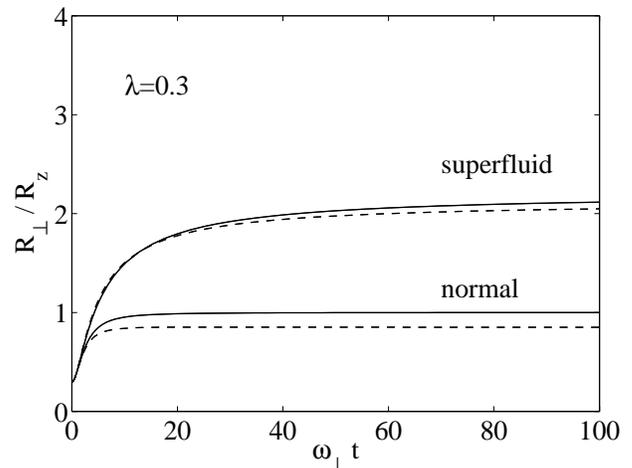}
\caption{Aspect ratio as a function of time for the expansion of the
normal (lower curves) and superfluid phase (upper curves) for $\lambda=0.3$
and  different values of the parameters $\chi$: 
$\chi=0$ (full line) and  $\chi=-0.4$ (dashed line).}
\label{fig2}
\end{figure}
\end{center}

The results of the numeric integration of the  equations of 
motion (\ref{intnormal}) are reported in Fig.\ref{fig1} and \ref{fig2}
as a function of time for the choices $\chi=0$ and $\chi=-0.4$.

We address now the problem of the expansion of a superfluid
Fermi gas. As already anticipated we will make use of  the hydrodynamic 
equations of superfluids. Those equations have been already used
to describe the collective oscillations of a superluid trapped Fermi 
gas \cite{petrov-baranov} including its rotational behaviour 
\cite{minguzzi_zambelli}.   The hydrodynamic equations 
are  applicable if the healing length is much smaller than the size 
of the sample, which implies that the energy gap should be larger 
than the oscillator energies $\hbar \omega_z$, $\hbar \omega_{\perp}$. 
This non trivial condition implies that the whole system behaves like
superfluid.
Furthermore the hydrodynamic equations are applicable
up to excitation energies of the order of the  energy gap. 
In the problem of the expansion it is crucial that the system 
remains superfluid 
in the first instants when the hydrodynamic forces provide the 
relevant acceleration to the expanding atoms. One expects that this 
condition be satisfied if the initial temperature  is  small enough. 

The hydrodynamic description is based on the equation of continuity

\begin{eqnarray}
\label{cont}
&& \frac{\partial}{\partial t}n+\nabla(n{\bf u})=0 
\end{eqnarray}
and on the Euler equation 

\begin{eqnarray}
\label{eul}
&& m\frac{\partial}{\partial t}{\bf u}+\nabla
\left( \mu_{\ell e}(n) +V_{ho}({\bf r})+\frac{1}{2}mu^2
\right)=0
\end{eqnarray}
where $\mu_{\ell e}(n)$ is the chemical potential of a uniform gas 
calculated at the density $n$ and ${\bf u}$ is the velocity field.

If the equation of state is a power law ($\mu_{\ell e} \propto n^{\gamma}$)
these  equations admit the simple scaling solution

\begin{eqnarray}
\label{scaling1}
n(r_i,t) &=& \frac{1}{\prod_j b_j} n_0\left(\frac{r_i}{b_i}\right) , \\
u_i(r_i,t) &=& \frac{{\dot b}_i}{b_i} r_i ,
\label{scaling2}
\end{eqnarray}
and the Thomas-Fermi radii evolve according to the law $R_i(t) = R_i(0) b_i(t)$.
In this case, it is immediate to show that,  during the expansion,
the scaling parameters obey the coupled differential equations 

\begin{equation}
\ddot{b}_i=\frac{\omega_i^2}{(b_xb_yb_z)^\gamma b_i}.
\label{bigamma}
\end{equation}
which reduce to  

\begin{equation}
\ddot{b}_z = \frac{\omega_z^2}{b_\perp^{2\gamma}},
%
\hspace{0.25cm}
{\rm and}
\hspace{0.25cm}
%
\ddot{b}_\perp = \frac{\omega_\perp^2}{b_\perp^{(2\gamma+1)}}.
\label{exprad}
\end{equation}
for highly elongated configurations ($\lambda \ll 1$). For $\gamma=1$  
(corresponding to a Bose-Einstein condensed gas) 
 the equation for the radial motion
is integrable analytically and one finds the result 
$b_{\perp}(t) = (1+\omega_\perp^2t^2)^{1/2}$ \cite{castin}.

To describe the expansion of superfluid Fermi gas we will use the same 
equation of state (\ref{mugg}) as for the normal phase. 
The case of a very dilute gas is also described by a power law 
with $\gamma =2/3$ (first term in (\ref{mugg})). 
For $\lambda=0.1$ and $0.3$ the solution is given by the full upper line in
Fig.\ref{fig1} and \ref{fig2} respectively which show that the 
deformation of the trap is inverted in time and the aspect ratio 
$R_{\perp}/R_z$ reaches asymptotically a value significantly larger  
than 1 \cite{gora1}.
Superfluidity has hence the effect of distributing the release energy
in a strongly asymmetric way along the axial and radial directions. 
It is worth noticing that the same scaling equations (\ref{bigamma}),
with $\gamma=2/3$, are obtained for a classic gas in the collisional regime
\cite{gora2}.

In the more general case (\ref{mugg}), a useful 
approximation to the solution of the hydrodynamic equations, 
based on the  scaling ansatz (\ref{scaling1},\ref{scaling2}), is obtained
by  multiplying 
the Euler's equation (\ref{eul}) by $r_i n({\bf r})$ and integrating
over the spatial coordinates. 
Using the equation of state (\ref{mugg}), one  finally obtains the following 
set of differential equations

\begin{eqnarray}
&& {\ddot b}_i + \omega_i^2 b_i 
- \frac{\omega_i^2}{b_i} \frac{1}{(\prod_i b_i)^{2/3}} +
 \nonumber \\
&& + \frac{3}{2}\chi \frac{\omega_i^2}{b_i}
\left(
\frac{1}{(\prod_i b_i)^{2/3}} - \frac{1}{\prod_i b_i} \right) 
=0,
\label{bgg}
\end{eqnarray}
with  $\chi$ defined by Eq.(\ref{chi}). Equations (\ref{bgg}) differ from the anlalogous
equations (\ref{intnormal}) holding in the normal phase. 
By linearizing Eqs.(\ref{bgg}) around $b_i=1$ one gets, in the 
case of a spherical trap, the result  $\omega_Q = \sqrt{2} \omega_0$ 
for the quadrupole frequency \cite{petrov-baranov}, which, contrary
to (\ref{omega_vichi}),  is independent of the 
interaction term in $\chi$.

The predictions of  Eqs.(\ref{bgg}) for the expansion of the gas   
are reported in Figs.\ref{fig1},\ref{fig2} and show that the  
inclusion of the interaction term in the equation of state affects 
the expansion of the superfluid only in a minor way.

In conclusion we have shown that the expansion of a superfluid Fermi
gas, being governed by the equations of hydrodynamics, differs in a
crucial way from the one of a normal gas in the collisionless regime. From
a theoretical point of view several questions remain to be investigated:
among them, the effect of large scattering lengths \cite{holland}
on the equation of state and the role of collisions which, under certain 
conditions, might give rise to a hydrodynamic regime, and hence to 
anisotropic expansion, also in the normal phase. Finally one should develop 
the formalism at finite temperature where both the normal and superfluid
components are present. The resulting bimodal structure in the expanding
cloud is expected to be affected by the transfer of atoms from the
superfluid to the normal component during the first stage of the
expansion.

This research is supported by the Ministero dell'Istru-\\zione, 
dell'Universit\`a e della Ricerca (MIUR). 
We acknowledge R.~Combescot, G.~Shlyapnikov and J.~E.~Thomas for useful 
discussions.




\end{document}